\documentstyle[12pt,twoside,fleqn,espcrc1,epsf]{article}
\newcommand{\pbp}{\langle {\bar \psi} \psi \rangle}
\newcommand{\Dslash}{$D$\kern-0.6em \hbox{/}}

\newcommand{\la}{\raise.16ex\hbox{$\langle$}}
\newcommand{\ra}{\raise.16ex\hbox{$\rangle$}}
\newcommand{\lap}{\mbox{}_{\textstyle \sim}^{\textstyle < }}

\newcommand{\be}{\begin{equation}}
\newcommand{\ee}{\end{equation}}
\newcommand{\beq}{\begin{eqnarray}}
\newcommand{\eeq}{\end{eqnarray}}

\title{The Role of Center Vortices in QCD\thanks{Talk presented by C. Alexandrou}}

\author{
C. Alexandrou\address{PSI, CH-5232
 Villigen, Switzerland and University of
 Cyprus, CY-1678 Nicosia, Cyprus},
Ph. de Forcrand\address{ETH, CH-8092 Z\"urich, Switzerland} 
and M. D'Elia\address{University of Pisa and INFN, I-56127 Pisa, Italy} }

\begin{document}
\maketitle

\begin{abstract}
Center vortices are unambiguously identified after Laplacian Center
Gauge fixing and their influence on confinement and chiral symmetry
breaking is investigated on a sample of SU(2) configurations at zero
and finite temperature.
\end{abstract}

\section{Introduction}

Within the vortex theory of confinement, put forward about twenty years ago 
\cite{history}, the QCD vacuum is considered as a condensate of colour
magnetic vortices with a flux quantized in terms of the center of the
group $Z_N$. A center vortex is associated with a singular gauge 
transformation with a discontinuity given by a gauge group
center element. It has the effect of multiplying the Wilson loop  linked to
this vortex by an element of $Z_N$, i.e. $W(C) \rightarrow e^{2\pi n
  i/N} W(C)$, $ n=1,...,N-1$. Assuming that center vortices 
condense in the QCD vacuum, the area law behaviour of large Wilson
loops follows from fluctuations in the number of vortices
linking the loops.

A procedure to identify these vortices on the lattice via gauge fixing
was shown recently to yield results in accord with  vortex
condensation theory~\cite{Greensite}. The main idea consists of
choosing a gauge that makes the link variables $U$ as close as
possible to the center of the gauge group. The Direct Maximal Center (DMC)
gauge proposed by Del Debbio {\it et al.}~\cite{Debbio} determines a
gauge transformation 
$g \in {\rm  SU(N)}$ that maximises the quantity
\be
 R[U] = \sum_{x,\mu} |{\rm Tr}~U^{GF}_\mu(x)|^2
\label{DMC}
\ee 
where $U^{GF}_\mu(x) = g(x)U_\mu(x)g^{\dagger}(x+\hat{\mu})$.
Then the gauge fixed links are projected onto $Z_N$, i.e. one replaces
each $U^{GF}$ by its closest center element $Z$ in the evaluation of the observables. 
For SU(2) the center projection is defined by
\be
Z_\mu(x) = {\rm sign} \>\>\left[{\rm Tr}~U^{GF}_\mu(x) \right]{\bf  I} 
\ee
and from now on, for simplicity, we will be discussing only SU(2).
 A plaquette in the $Z_2$-projected theory with value  $-1$
represents a defect called a P-vortex.
 The idea of center dominance is that center vortices, identified as
P-vortices in the DMC gauge,  are the relevant 
 nonperturbative degrees of freedom. It is supported by numerical results:
the $Z_2$-theory shows a string tension similar to that 
of the full- theory~\cite{Debbio};
the deconfinement phase transition~\cite{Engelhardt}
is well described  in the center vortex picture as a percolation
transition. 

The center-dominance scenario received further support by the
following test carried out 
in Ref.~\cite{PRL}: The P-vortices were removed by considering 
a modified ensemble with links
$U'_\mu (x) \equiv Z_\mu(x) U_\mu(x)$
which  projects onto the trivial $Z_2$ vacuum. It was shown~\cite{PRL} that
in this modified ensemble,
both confinement is lost and chiral symmetry is restored.

We have investigated the chiral content of the $Z_2$-projected theory further,
by looking at the quark condensate $\pbp$ in the $Z_2$ sector 
as a function of the quark
mass $m_q$, at zero and finite temperature. At zero temperature 
it extrapolates linearly to a non-zero value as $m_q \rightarrow
0$ as shown in Fig.~1. For very small quark masses 
it diverges as $1/m_q$, revealing the
 presence of a few extremely
small eigenvalues, which may be caused by the non-trivial topology
of the original $SU(2)$ gauge field. 
This behaviour is strikingly similar to that observed with domain-wall fermions
~\cite{domain-wall}. It is well described by the functional form
\be
\pbp_{Z_2} = a + b/m_q + c \> m_q \quad.
\ee
This ansatz also works well at finite temperature.
The comparison of data on $16^3 \times N_t$ volumes at $\beta=2.4$, 
with $N_t=16, 8$ and $4$
(the latter in the deconfined phase) reveals that $a, b$ and $c$ do not 
vary much with temperature. $c$ remains close to its free-field value.
$a$ tends to decrease somewhat in the deconfined, chirally symmetric phase, 
but remains surprisingly large. $b$ may also show some decrease, but not in
the same proportion as the fluctuations of the $SU(2)$ topological charge.
Therefore, one has to be cautious in relating the $Z_2$ condensate to the
chiral properties of $SU(2)$.

\begin{figure}[!h]
\begin{center}
\epsfxsize=6.5cm
\epsfysize=6.0cm
\epsffile[-257 195 271 669]{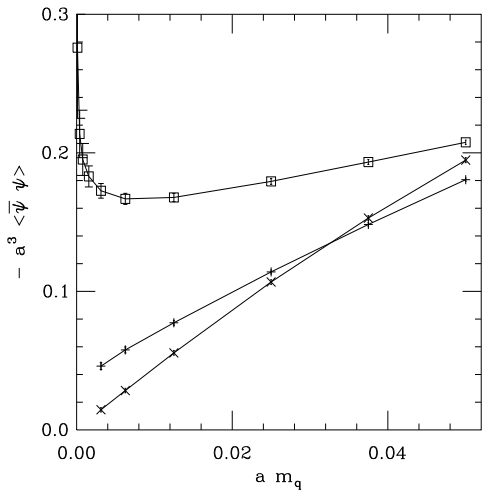}
\vspace*{-1.7cm}
\caption{Quark condensate versus quark mass, in the
  original (+),
the center-vortex free (x), and the center-projected (squares) theories.}
\end{center}
\vspace*{-2.cm}
\end{figure}

\section{Laplacian Center Gauge}

A local iterative maximization of the DMC gauge-fixing condition Eq.~\ref{DMC} 
selects any one of the many possible maxima.
Each of these Gribov copies will 
have its own set of P-vortices, which may show dramatically different 
properties~\cite{Kovacs}.
In order to find a  center gauge fixing
procedure  that is  free
of Gribov copies, we first note
that DMC is equivalent to maximizing 
$\sum_{x,\mu} {{\rm Tr_{adj}}}~U_\mu(x)$,
since 
\be
|{\rm Tr}~U|^2={\rm Tr}~U^2 +2  
= {\rm Tr_{adj}}~U + 1 \hspace*{0.5cm} {\rm taking }\hspace*{0.5cm} 
{\rm Tr}~\sigma_a \sigma_b = 2\delta_{ab}
\quad .
\ee 
The idea is thus to smooth the center-blind, adjoint component of the
gauge field as much as possible, then to read the center component off
the fundamental gauge field. Therefore, Maximal Center Gauge is just another
name for adjoint Landau gauge.

The problem of Gribov copies in the fundamental Landau gauge was solved
in \cite{VW}:
If one relaxes the requirement that $g \in SU(2)$,
the maximization of the gauge-fixing functional is achieved by taking for
$g^\dagger$ the eigenvector  $\vec{v}$ associated with the
smallest eigenvalue of the covariant Laplacian 
$\Delta_{xy} = 2d \>\delta_{xy} - \sum_{\pm \hat{\mu}} U_{\pm \hat{\mu}}(x)
\delta_{x \pm \hat{\mu},y}$.
 At each site,
$v(x)$ has 2 complex color components. The Laplacian gauge condition consists
of taking for $g^\dagger$ the $SU(2)$ projection of $v$, thus 
rotating $v(x)^\dagger$ along direction $(1,1)$ at all sites.

We  follow this construction for the adjoint representation. The covariant
Laplacian is now constructed from adjoint links
$U^{ab} = \frac{1}{2} \mbox{Tr}~[U \sigma^a U^\dagger \sigma^b],~a,b=1,2,3$.
It is a real symmetric matrix. The lowest-lying eigenvector $\vec{v}$ has
3 real components $v_i, \> i=1,2,3$
at each site $x$. One can apply a local gauge transformation $g(x)$ to rotate
it along some fixed direction.
Note, however, that this does not specify the gauge completely: Abelian
rotations around this reference direction are still possible.
What we have achieved at this stage is a variation of Maximal Abelian Gauge
which is free of Gribov ambiguities. This Laplacian Abelian Gauge has been
proposed in \cite{LAG} and, as it was shown there,  monopoles 
 are directly identifiable by the condition $|v(x)| = 0$
for smooth fields. Abelian monopole worldlines appear naturally as the locus
of ambiguities in the gauge-fixing procedure: the rotation to apply to 
$v(x)$ cannot be specified when $|v(x)| = 0$.

To fix to center gauge, we must go beyond Laplacian Abelian Gauge and specify
the Abelian rotation. This is done most naturally
by considering the second-lowest eigenvector $\vec{v'}$ of the adjoint
covariant Laplacian, and requiring that the plane $(v(x),v'(x))$ be parallel
to, for instance, $(\sigma_3,\sigma_1)$ at every site $x$. This fixes the gauge
completely, except where $v(x)$ and $v'(x)$ are collinear. Collinearity 
occurs when $\frac{v_1}{v'_1} = \frac{v_2}{v'_2} = \frac{v_3}{v'_3}$,
i.e. 2 constraints must be satisfied. Thus, gauge-fixing ambiguities have
codimension 2: in 4$d$, they are 2$d$ surfaces. They
can be considered as the center-vortex cores~\cite{lat99}.

\vspace*{-0.3cm}

\section{Results}

\vspace*{-0.3cm}

We have applied Laplacian Center Gauge fixing and center projection to an
ensemble of $SU(2)$ configurations. 
The main difference with DMC gauge is an increase in the $P$-vortex density 
($\sim 11\%$ vs $\sim 5.5\%$
on a $16^4$ lattice at $\beta=2.4$), similar to the increase in
the monopole density for Laplacian Abelian Gauge~\cite{LAG}.
As in \cite{PRL}, 
the string tension, the quark condensate and the 
topological charge all vanish upon removal of the $P$-vortices. 
Fig.~2 displays the Creutz ratios
$\chi(R,R) = -{\rm ln}\left[\langle W(R,R)\rangle \langle W(R-1,R-1)\rangle/\langle W(R,R-1)\rangle^2\right]$
constructed from averages $\langle W(R,T)\rangle$ of $R$ by $T$ Wilson loops. For
large $R$, $\chi(R,R)$ approaches the string tension $\sigma$.  On the
modified
configuration the string tension goes to zero whereas
 in the $Z_2$-projected theory it reproduces the $SU(2)$ value of Ref.~\cite{MT}.
The quark condensates behave  as in Fig.~1. 

\begin{figure}
\begin{minipage}{7cm}
\vspace*{-0.3cm}
\epsfxsize=6.truecm 
\epsfysize=6.truecm 
\mbox{\epsfbox{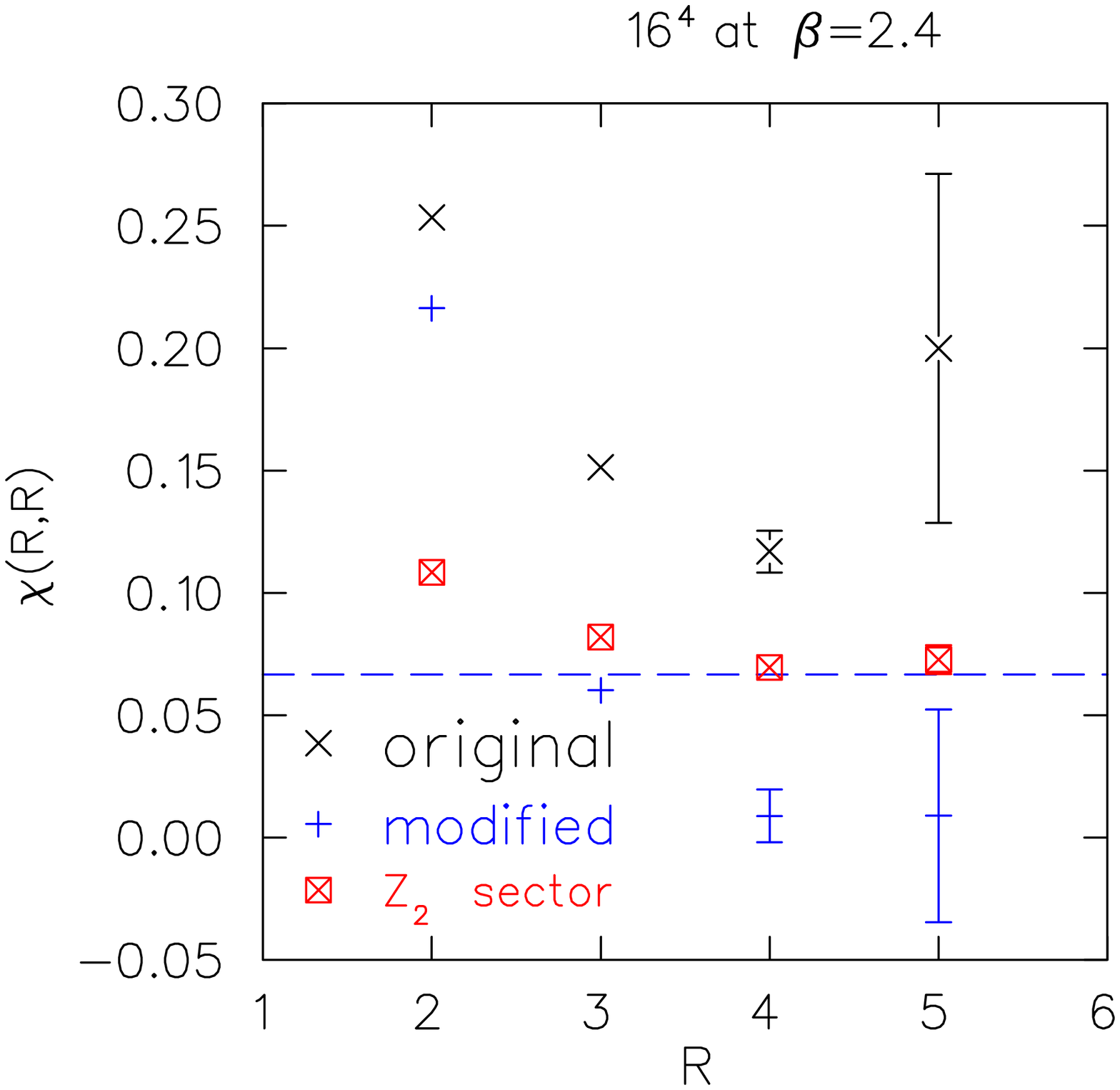}}
\vspace*{-0.9cm}
\caption{Creutz ratios for the original, the modified and the $Z_2$
  projected ensembles. The dashed line is the string tension result of 
  Ref.~10.}
\vspace*{-1.cm}
\end{minipage} \hfill
\begin{minipage}{8cm}
\vspace*{-2.6cm}
\epsfxsize=8.5truecm 
\epsfysize=8.5truecm
\epsffile{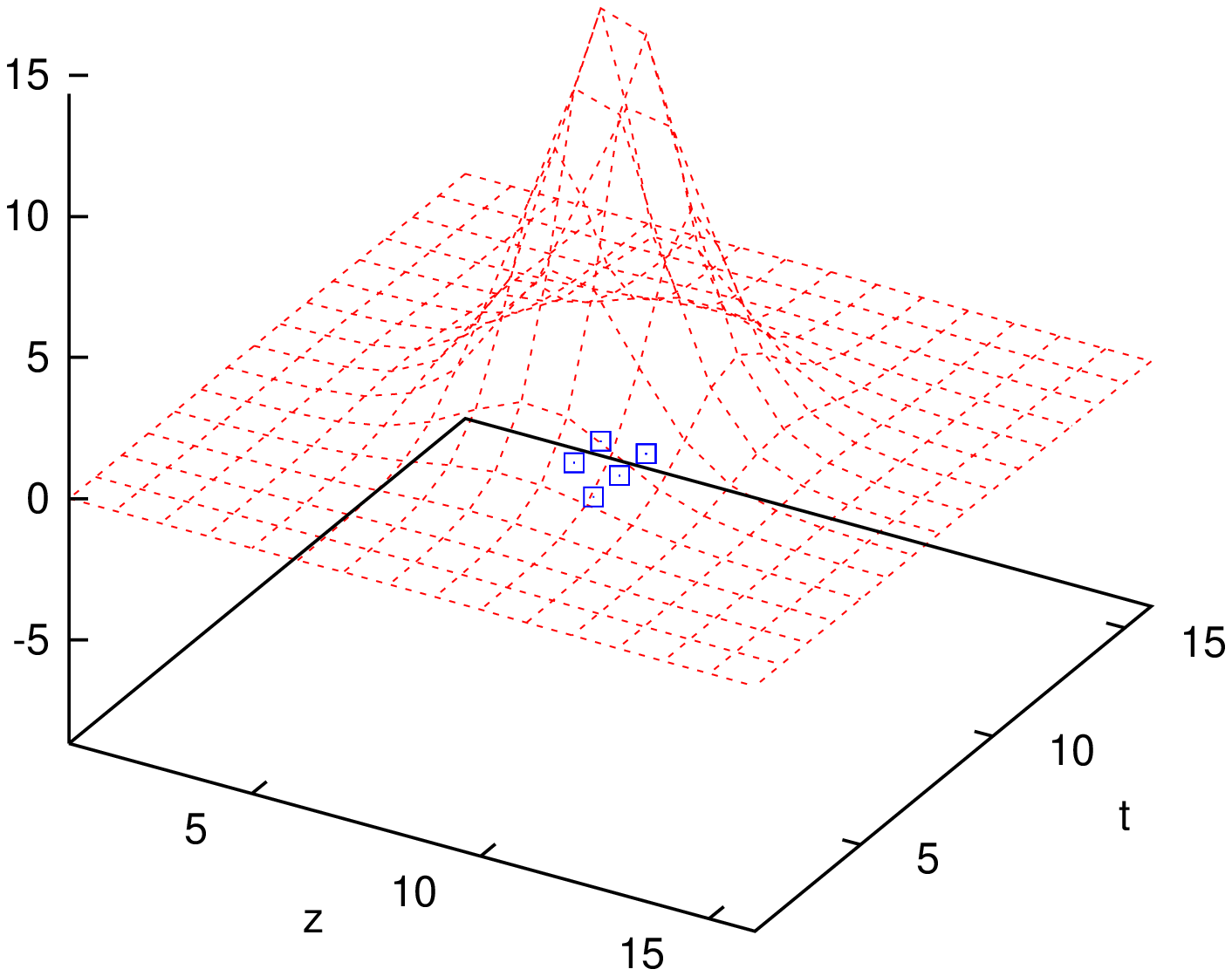}
\vspace*{-1.8cm}
\caption{Location of P-vortices on a cooled instanton configuration.
The surface shows the action density and the squares are the P-vortices.}
\vspace*{-0.9cm}
\end{minipage}
\end{figure}

If one applies Laplacian Center Gauge fixing to a cooled one-instanton
configuration, one finds a very small number ($\lap 100$) of P-vortices,
regardless of the original instanton size. These P-vortices are all 
 near the instanton center, as illustrated in Fig.~3.

\vspace*{-0.3cm}

\section{Conclusions}

\vspace*{-0.3cm}

Any Gribov ambiguities that cast doubt on the physical
    relevance of P-vortices (pointed out e.g. by ~\cite{Kovacs})
are removed by the Laplacian Center Gauge
    fixing. This gauge appears naturally as an extension of Laplacian Abelian
 Gauge. It allows the direct identification of center vortices (and monopoles) 
 by inspection of the two lowest eigenmodes of
 the covariant adjoint Laplacian, without gauge-fixing. 
As in DMC gauge, center dominance seems to hold: 
 (i) At zero temperature the string tension is well reproduced in the
$Z_2$-projected theory, and even the $Z_2$ quark condensate is non-zero;
(ii) In the deconfined phase, the $Z_2$ string tension vanishes: however the
$Z_2$ quark condensate does not.

Although here we have only discussed SU(2), 
 our gauge fixing procedure readily generalizes to $SU(N)$: complete gauge-fixing
is achieved by rotating the first $(N^2-2)$ eigenvectors of the adjoint
Laplacian along some reference directions. Ambiguities arise whenever
these $(N^2-2)$ eigenvectors 
[each with $(N^2-1)$ real components] 
become
linearly dependent. This again defines codimension-2 center-vortex cores.

\end{document}